\documentclass[twocolumn,showpacs,prl,letterpaper]{revtex4}

\usepackage{amsfonts,amssymb,amsmath,amsthm}
\usepackage{graphicx}
\usepackage{epsfig}

\newcommand{\ncd}{\newcommand}
\ncd{\QCcns}{$QC_{\cal{C}}$}
\ncd{\QCc}{$QC_{\cal{C}}\;$}

\begin{document}


\title{Fault-tolerant quantum computation with high threshold
  in two dimensions}

\author{$\mbox{Robert Raussendorf}^1$ and $\mbox{Jim Harrington}^2$}
\affiliation{1: Perimeter Institute for Theoretical Physics, 31 Caroline
  St. N., Waterloo, ON N2L 2Y5, Canada\\
2: Applied Modern Physics, MS D454, Los Alamos National Laboratory, 
Los Alamos, NM 87545, USA}

\begin{abstract}
   We present a scheme of fault-tolerant quantum computation for a
    local architecture in two spatial dimensions. The error threshold
    is  0.75\% for each source in an error model
    with preparation, gate, storage and measurement errors.
\end{abstract}


\maketitle

Quantum computation is fragile. Exotic quantum states are created in
the process, exhibiting entanglement among large numbers of particles
across macroscopic distances. In realistic physical systems, decoherence 
acts to transform these states into more classical ones, compromising their
computational power. Fortunately, the effects of decoherence can be
counteracted by quantum error correction \cite{ShorE}. In fact, arbitrarily 
large quantum computations can be performed with arbitrary accuracy,
provided the error level of the elementary components of the quantum 
computer is below a certain threshold. This is guaranteed by the threshold 
theorem for quantum computation \cite{TT1,TT2,TT3,TT4}.

Now that the threshold theorem has been established, it is important 
to devise methods for error correction which yield a high threshold, 
are robust against variations of the error model, 
and can be implemented with small operational overhead. 
An additional desideratum is a simple architecture for the quantum 
computer, requiring no long-range interaction, for example.

Recently, a threshold estimate of $3\times 10^{-2}$ per operation
has been obtained for a method using post-selection \cite{Kn2}. 
An alternative scheme with high threshold combines topological 
quantum computation with state purification \cite{BT}. (See also 
\cite{BMD}.) In that approach, a subset of the universal gates are 
assumed to be error-free. 
Pure topological quantum computation ideally requires no
error correction but often picks up a comparable poly-logarithmic
overhead \cite{SKoh} in the Solovay-Kitaev construction for
approximating single- and two-qubit gates (c.f. \cite{Bon}).
fault tolerance is more difficult to achieve in architectures where
each qubit can only interact with other qubits in its immediate
neighborhood. A fault tolerance threshold for 
a two-dimensional lattice of qubits with only local and
nearest-neighbor gates is $1.9 \times 10^{-5}$ \cite{Svo}.

In this Letter, we present a scheme for fault-tolerant universal
quantum computation on a two-dimensional lattice of
qubits, requiring only a nearest-neighbor translation-invariant
Ising interaction and single-qubit preparation and measurement. 
A fault tolerance threshold of $7.5 \times 10^{-3}$ for each error 
source is presented, with moderate resource scaling. This scheme is best 
suited for implementation with massive qubits where geometric 
constraints naturally play a role, such as cold atoms in optical
lattices \cite{Ola} or two-dimensional ion traps \cite{Hen}.

\begin{figure}[t]
  \begin{center}
  \epsfig{width=8.6cm,file=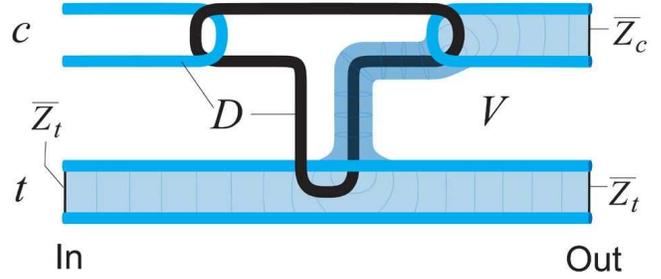}
  \caption{\label{CNOThomol} (Color online.) The CNOT gate 
    $\Lambda(X)_{c,t}$ ($c$: control, $t$: target) formed by topologically
    entangled lattice defects. Each pair of defects carries an encoded
    qubit. Defects exist as primal (blue) and dual (black), and are
    created by local  measurement. The primal correlation surface (light blue)
    shown here converts an incoming Pauli operator
    $Z_t$ into an outgoing $Z_t\otimes Z_c$, as required for a 
    CNOT gate. }
  \end{center}
\end{figure}

The presented scheme integrates methods of topological quantum
computation, specifically the toric code \cite{Kit1}, and magic
state distillation \cite{BK04} into the one-way quantum computer
(\QCcns) \cite{RBB03} on cluster states. By employing magic state
distillation we improve the error threshold significantly
beyond \cite{RHG}, with the threshold value and overhead scaling now set
by the topological error correction. In this regard, we would like
to emphasize that the three-dimensional cluster state is an
intrinsically fault-tolerant substrate for quantum computation
\cite{RHG}. From the viewpoint of implementation it is desirable to
reduce the spatial dimensionality of the scheme from three to two. To
achieve this we 
turn the \QCc into a sequential scheme in which the cluster state is
created slice by slice. 

This Letter is organized as follows. First, we construct
fault-tolerant universal gates for the \QCc in three spatial
dimensions.  (See Fig.~\ref{CNOThomol} for a CNOT gate.)  
Next, we perform the mapping to two dimensions. Finally, we present
our error model and work out its threshold value.
 
\begin{figure}[t]
  \begin{center}
  \epsfig{width=8.6cm,file=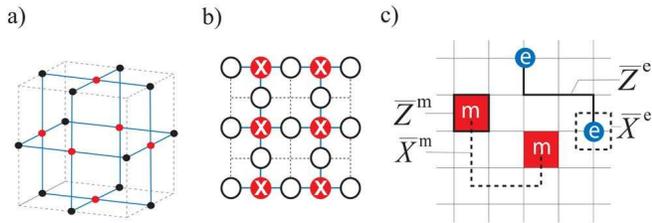}
  \caption{\label{LattDef}(Color online.) Lattice definitions. a)
    Elementary cell of the cluster lattice ${\cal{L}}$. 1-chains of
    ${\cal{L}}$ (dashed lines), and graph edges (solid lines). b) A
    surface code obtained from a 2D cluster state by local
    $X$-measurements. c) A pair of electric (``e'') or magnetic
    (``m'') holes in the code
    plane each support an encoded qubit. $\overline{Z}^{e/m}$
    and $\overline{X}^{e/m}$ denote the encoded Pauli operators $Z$ and
    $X$, respectively.}
  \end{center}
\end{figure}

We consider a cluster state
$|\phi\rangle_{\cal{L}}$ on a lattice ${\cal{L}}$ with elementary cell 
as displayed in Fig.~\ref{LattDef}a. Qubits are located at the center 
of faces and edges of ${\cal{L}}$. The lattice ${\cal{L}}$ is subdivided 
into three regions $V$, $D$ and $S$.  Each region has its purpose, 
shape and specific measurement basis for its qubits. The qubits in $V$ 
are measured in the $X$-basis, the qubits in $D$ in the $Z$-basis, 
and the qubits in $S$ in either the $Y$-basis or the eigenbasis 
$(X + Y)/\sqrt{2}$.  $V$ fills up most of the cluster. $D$ is composed 
of thick line-like structures, named {\it defects}. $S$ is composed of 
well-separated qubit locations interspersed among the defects.  As 
described in greater detail below, the cluster region $V$ provides 
topological error correction, while regions $D$ and $S$ specify the 
Clifford and non-Clifford parts of a quantum algorithm, respectively.

We can break up this measurement pattern into gate simulations by 
establishing the following correspondence:
$\mbox{\em{quantum gates}} \leftrightarrow 
\mbox{\em{quantum correlations}} \leftrightarrow \mbox{\em{surfaces}}$, 
as illustrated for the CNOT gate in Fig.~\ref{CNOThomol}. 
The first part of this correspondence has been established in 
\cite{RBB03}. For the second part homology comes into play. 
The correlations of $|\phi\rangle_{\cal{L}}$ (i.e., the stabilizers) can 
be identified with 2-chains (surfaces) in ${\cal{L}}$, while errors map 
to 1-chains (lines). Homological equivalence of the chains implies 
physical equivalence of the corresponding operators \cite{RHG}. 
This correspondence is key to the presented scheme. Gates are 
specified by a set of surfaces with input and  output boundaries, 
and syndrome measurements correspond to closed surfaces 
(having no boundary).

Formally, ${\cal{L}}$ is regarded as a chain complex,
${\cal{L}}=\{C_3,C_2,C_1,C_0\}$. It has a dual
${\overline{\cal{L}}}=\{\overline{C}_3,\overline{C}_2,\overline{C}_1,
\overline{C}_0\}$ whose cubes $\overline{c}_3 \in \overline{C}_3$
map to sites $c_0 \in C_0$ of ${\cal{L}}$, whose faces
$\overline{c}_2 \in \overline{C}_2$ map to edges $c_1 \in C_1$ of
${\cal{L}}$, etc. The chains have coefficients in $\mathbb{Z}_2$. One
may switch back and forth between
${\cal{L}}$ and $\overline{\cal{L}}$ by a duality transformation
$\mbox{}^*(\;)$.  ${\cal{L}}$, $\overline{\cal{L}}$ are equipped with
a boundary map $\partial$, where $\partial \circ \partial = 0$.

Operators may be associated with chains as follows. Suppose that for each
qubit location $a$ in a chain $c$, $a \in \{c\}$, there exists an operator
$\Sigma_a$, with $[\Sigma_a,\Sigma_b]=0$ for all $a,b \in
\{c\}$. Then, we define $\Sigma(c) := \prod_{a \in \{c\}} \Sigma_a$.
Cluster state correlations are associated with
2-chains. Specifically, all
elements in the cluster state stabilizer take the form
$K(c_2)K(\overline{c}_2)$ with $c_2 \in C_2$, $\overline{c}_2 \in
\overline{C}_2$, and
\begin{math}
  K(c_2)= X(c_2)Z(\partial c_2),\;\; K(\overline{c}_2)=
  X(\overline{c}_2)Z(\partial \overline{c}_2).
\end{math}
Only those stabilizer elements compatible with the local measurement
scheme are useful for information processing. In particular, they need to
commute with the measurements in $V$ and $D$,
\begin{equation}
  \label{CompatCond}
  \begin{array}{rclcr}
    {[K(c_2)K(\overline{c}_2), X_a]}  &=& 0, && a \in V, \\
    {[K(c_2)K(\overline{c}_2), Z_b]}  &=& 0, && b \in D.
  \end{array}
\end{equation}
This condition may again be expressed in terms of the chains $c_2$,
$\overline{c}_2$ directly, which we will do below.

\paragraph{Topological error correction in $V$.}  
Inside $V$ the constraint (\ref{CompatCond}) implies $\partial c_2=0$, 
$\partial \overline{c}_2=0$. In particular, these conditions are obeyed 
for $c_2 = \partial c_3$, $\overline{c}_2 = \partial \overline{c}_3$. 
For each elementary cube $q \in C_3$, $\overline{q} \in \overline{C}_3$ 
the cluster stabilizers $K(\partial q)$, $K(\partial \overline{q})$ can be 
measured by the local $X$-measurement and classical post-processing. 

The optimal error correction procedure for $V$ can be mapped to a
model from classical statistical mechanics, the {\em{random plaquette 
$\mathbb{Z}_2$-gauge model in three dimensions}} (3D-RPGM) \cite{DKLP}, 
for which a fault tolerance threshold of $3.3\times 10^{-2}$ for local noise 
has been found in numerical simulations \cite{Ohno}. (See also \cite{TSN}.) 
Here we use the minimum weight chain matching algorithm \cite{AlgoEff} 
for error correction. It yields a slightly smaller threshold of $2.9
\times 10^{-2}$  
\cite{Harri} but is computationally efficient. Various error sources eat away 
at this 3\% error budget. 

\paragraph{Cluster states and surface codes.} 
The connection between a 2D cluster state and a surface code is
illustrated in Fig.~\ref{LattDef}b. The extra spatial dimension in
a 3D cluster state allows to evolve coded states in ``simulated time''.
The number of qubits which can be encoded in a surface code depends 
solely on the surface topology. Here we consider a plane with pairs of 
either electric or magnetic holes; see Fig.~\ref{LattDef}c.  A magnetic 
hole is a plaquette $f$ where the associated stabilizer generator 
$S_\Box(f)=Z(\partial{f})$ is {\em{not enforced}} on the code space, 
and an electric hole is a site $s$ where the associated stabilizer 
$S_+(s)=X(\partial \mbox{ }^\# s)$ is {\em{not enforced}} on the code 
space, where ``$\#$'' denotes the duality transformation in 2D.  Each 
hole is the intersection of a defect strand with a constant-time slice.

A pair of holes supports a qubit. For a pair of magnetic holes $f,
f'$, the encoded spin flip operator is $\overline{X}^m =
X(\overline{c}_1)$, with $\{\partial \overline{c}_1\} =\{\mbox{}^\#f,
\mbox{}^\#f^\prime \}$, and the
encoded phase flip operator is $\overline{Z}^m= Z(c_1)$, with $c_1
\cong \partial f$ or $c_1 \cong \partial f^\prime$. The operator
$Z(\partial f + \partial f^\prime)$ is in the code stabilizer. For a
pair of electric holes $s, s^\prime$ we have $\overline{X}^e =
X(\overline{c}_1^\prime)$, with $\overline{c}_1^\prime \cong \partial
\mbox{}^\#s$, $\overline{Z}^e = Z(c_1)$, with $\{\partial c_1\} = \{s,
s^\prime \}$, and $X(\partial \mbox{}^\#s+ \partial
\mbox{}^\#s^\prime)$ is in the code stabilizer.

\paragraph{Quantum logic.} The CNOT gate is realized by linking 
primal and dual defects as displayed in Fig.~\ref{CNOThomol}. 
To explain the functioning of the gate we refer to Theorem 1 of 
\cite{RBB03}.  We consider a block shaped cluster ${\cal{C}}$ 
where the elementary cell of Fig.~\ref{LattDef}a is repeated an 
integer number of times along each direction. One of these 
directions is singled out as ``simulated time''. The two perpendicular 
slices of the cluster at the earliest and latest times contain the 
supports $I$ and $O$ for the encoded input and output qubits, 
respectively, with $I,O \subset \{C_1\}$ encoded by the surface
code of Fig.~\ref{LattDef}c.

The set $M$ on which the measurement pattern is defined (c.f. Thm~1 of
\cite{RBB03}) is composed of $V$ and $D$, $M=V \cup D$. Due to the
presence of a primal lattice ${\cal{L}}$ and a dual lattice
$\overline{\cal{L}}$, it is convenient to subdivide the sets $V$ and $D$
into primal and dual subsets. Specifically,
$V = V_p \cup V_d$, with $V_p \subset \{C_2\}$, $V_d \subset
\{\overline{C}_2\}$, and $D = D_p \cup D_d$, with $D_p \subset
\{C_1\}$, $D_d \subset \{\overline{C}_1\}$.

\begin{figure}[t]
  \begin{center}
  \epsfig{width=8.6cm,file=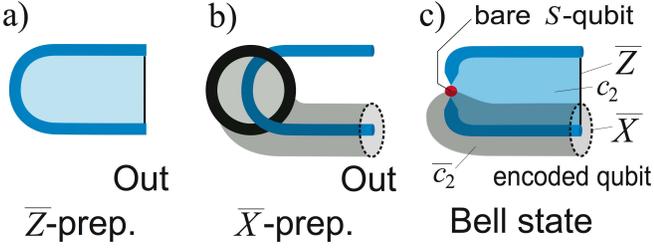}
  \caption{\label{gates}(Color online.) Remaining gates for universal
    fault-tolerant  
  computation.  The relevant correlation surfaces are shown in light
  blue and gray.  
     Replace Out(put) by In(put) for a measurement.
   a) Preparation of a $\overline{Z}$-eigenstate for an electric qubit.  
    b) Preparation of an $\overline{X}$-eigenstate for an electric qubit.  
    c) Creation of a Bell pair among a bare $S$-qubit and an encoded qubit.}
  \end{center}
\end{figure}

With these definitions, we can now prove the functioning of the
CNOT gate in Fig.~\ref{CNOThomol}. The gate cluster
${\cal{C}}$ contains the regions $V_p$, $V_d$, $D_p$, $D_d$, 
$I$ and $O$. In this setting, condition (\ref{CompatCond}) implies 
for the correlation surfaces:
\begin{equation}
  \begin{array}{rclcrcl}
    \{c_2\} &\subset& V_p, && \{\partial c_2\} &\subset& D_p \cup I
    \cup O,\\
    \{\overline{c}_2\} &\subset& V_d \cup I \cup O, && \{\partial
    \overline{c}_2 \} &\subset& D_d.
  \end{array}
\end{equation}
One such (primal) correlation surface is depicted in Fig.~\ref{CNOThomol}. 
The corresponding stabilizer of $|\phi\rangle_{\cal{C}}$, after measurement 
of the qubits in $M=V \cup D$, implies a stabilizer
$\pm\overline{Z}^e_{t,I} \overline{Z}^e_{c,O}
\overline{Z}^e_{t,O} = \pm \overline{Z}^e_{t,I} \Lambda(X)_{c,t}
\overline{Z}^e_{t,O} \Lambda(X)_{c,t}^\dagger$ for
$|\Psi\rangle_{IO}$. Three similar
surfaces imply the stabilizer elements $\pm \overline{X}^e_{t,I}
\overline{X}^e_{t,O}$, $\pm \overline{Z}^e_{c,I}
\overline{Z}^e_{c,O}$ and $\pm\overline{X}^e_{c,I} \overline{X}^e_{c,O}
\overline{X}^e_{t,O}$ for  $|\Psi\rangle_{IO}$.
Theorem~1 of \cite{RBB03} is applied with
$U=\Lambda(X)_{c,t}$. \qed

Further elements of a fault-tolerant
\QCcns-computation are shown in Fig.~\ref{gates}. 
Fault-tolerant preparation of encoded $X$- and $Z$-eigenstates
for the electric qubits are displayed in Figs.~\ref{gates}a and \ref{gates}b,
which can be reversed to denote measurements. These operations, 
together with the CNOT gate of Fig.~\ref{CNOThomol}, comprise the set 
of topologically protected gates. 
Fig.~\ref{gates}c shows the creation of a Bell pair between a bare
$S$-qubit and a qubit encoded with a surface code
(electric). The shown correlation surfaces $c_2$, $\overline{c}_2$ are
such that  $\{c_2\} \subset V_p$, $\{\partial c_2\} \subset D_p \cup S
\cup O$, $\{\overline{c}_2\} \subset V_d \cup S \cup O$, $\{ \partial
\overline{c}_2\}= \emptyset$. The corresponding stabilizers $K(c_2)$,
$K(\overline{c}_2)$ imply, after local measurement of the qubits in
$V$ and $D$, the stabilizer generators $\pm Z_S \overline{Z}_O$,  $\pm X_S
\overline{X}_O$ for the state $|\Psi\rangle_{SO}$. Thus,
$|\Psi\rangle_{SO}$ is a Bell state with the qubit located on
$O$ being encoded. Measurement of the bare qubit on $S$ in the
eigenbasis of $Y$ or $(X+Y)/\sqrt{2}$ yields on $O$ an encoded state
$|\overline{Y}\rangle=|\overline{0}\rangle+i|\overline{1}\rangle$
or $|\overline{A}\rangle=|\overline{0}\rangle+e^{i\pi/4}
|\overline{1}\rangle$, respectively. These states are noisy and
therefore subsequently purified via magic
state distillation \cite{BK04}. Finally, they are used in
teleportation circuits (see Fig.~10.25 of \cite{NC00}) to
generate the fault-tolerant gates $\exp(i\pi/4\, \overline{X})$ and 
$\exp(i\pi/8\, \overline{Z})$. This completes the universal
fault-tolerant gate set.

\paragraph{Mapping to the 2D lattice.} The dimensionality of the
spatial layout can be reduced by one if the cluster is created slice
by slice. That is, we convert
the axis of ``simulated time''---introduced as a means to explain the
connection with surface codes---into real time. 

Cluster qubits located on time-like edges of ${\cal{L}}$ or
$\overline{\cal{L}}$ 
become syndrome qubits, which are periodically measured.  Qubits on
space-like edges become code qubits.  Time-like oriented $\Lambda(Z)$ gates 
are mapped to Hadamard gates, while space-like oriented $\Lambda(Z)$
gates remain unchanged. 

\begin{figure}[t]
  \begin{center}
  \epsfig{height=3.1cm,file=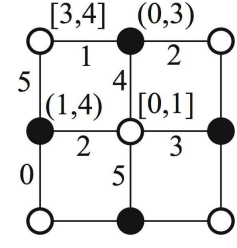}
  \caption{\label{Cell2D}Elementary cell of the 2D
    lattice. Temporal order of operations in $V$: The labels on the edges
    denote the time steps at which the corresponding $\Lambda(Z)$ gate 
    is performed. The labels at
    the syndrome vertices (``$\circ$'') denote measurement and
  (re-)preparation times 
    $[t_M, t_P]$, and the labels at the code vertices (``$\bullet$'')
  denote times for Hadamard gates $(t_H,t_H^\prime)$. The pattern is periodic
  in space, and in time with period six.}
  \end{center}
\end{figure}

The temporal order of operations is displayed in
Fig.~\ref{Cell2D}. Note that every qubit is acted upon by an operation in every
time step. The mapping to the two-dimensional structure has no
impact on information processing. In particular, the
error correction procedure is still the same as in fault-tolerant
quantum memory with the toric code.

\paragraph{Error model and threshold.} There are two separate thresholds, 
one for the Clifford operations and one for the non-Clifford operations. 
The former threshold derives from topological error correction 
and the latter from magic state distillation. The overall threshold 
is set by the smaller of the two. 

Mapping to a single-layer 2D structure  slightly modifies the
effective error model on the 
lattices ${\cal{L}}$ and $\overline{\cal{L}}$, as compared to \cite{RHG}.
Specifically, we assume the following:
1) Erroneous operations are
modeled by perfect operations preceded or followed by a partially depolarizing 
single- or two-qubit error channel $T_1 = (1-p_1)[I]+p_1/3\,([X]+[Y]+[Z])$, 
$T_2 = (1-p_2)[I]+p_2/15\,([X_aX_b]+..+[Z_aZ_b])$.  The error
sources are a) the preparation of the individual qubit states $|+\rangle$ 
(error probability $p_P$), b) the Hadamard gates (error
probability $p_1$), c) the $\Lambda(Z)$ gates (error
probability $p_2$),  
d) measurement (error probability $p_M$). 
2) Classical syndrome processing is instantaneous. 

When calculating a threshold, we assume that 
all error sources are equally strong, $p_1=p_2=p_M=p_P:=p$. Storage
errors need  not be considered because no qubit is ever idle between
preparation and measurement.  
This model encompasses realistic error sources such as local
inhomogeneity of electric and magnetic fields, fluctuations in laser intensity,
and imperfect photodetectors.  

The topological threshold for each physical source is estimated by 
numerical simulations to be
\begin{equation} \label{TopT}
  p_c = 7.5\times 10^{-3}.
\end{equation}
A similar threshold persists under modifications of the error model
such as higher weight errors \cite{RHG}.

Regarding the distillation threshold, the residual error 
$\epsilon_l$ at level $l$ undergoes the recursion 
$\epsilon_l \rightarrow \epsilon_{l+1} = 35 {\epsilon_l}^3$  
(to leading order) \cite{BK04}. The initial distillation error
$\epsilon_0$ arises through the effective error on an 
$S$-qubit, with $\epsilon_0 = 6p$.  The 
distillation threshold $p_c$ for each physical error source
is then $p_c = 1/6\sqrt{35} \approx 2.8\times 10^{-2}$. 
The purification threshold is much larger than the 
topological threshold, and therefore the overall threshold for
fault-tolerant \QCcns-computation is given by Eq.~(\ref{TopT}).

\paragraph{Overhead.} fault tolerance leads to a poly-logarithmic increase 
of operational resources. Both the overheads in topological error correction 
and in magic state distillation are described by a characteristic exponent: 
$\gamma_{top}=3$ and $\gamma_{ms}=\log_3 15$. The larger one 
dominates the resource scaling. Given bare circuit size $S$, the encoded 
circuit size $S^\prime$ scales as $S^\prime \sim S \log^3 S$. 

\paragraph{Conclusion.}
We have presented a scheme of fault-tolerant quantum computation in a 
two-dimensional local architecture with high error threshold and moderate 
overhead in resource scaling. The threshold of $7.5 \times 10^{-3}$ is the 
highest known for a local architecture. Our scheme only requires local and 
translation-invariant nearest-neighbor interaction in a single-layer
two-dimensional lattice.  
Small-scale experimental devices may be realized in optical lattices, 
segmented ion traps, or arrays of quantum dots or superconducting qubits 
where short-range interaction is preferred. 

\acknowledgments{
We would like to thank Frank Verstraete, Sergey Bravyi, Kovid Goyal, 
and John Chiaverini for helpful discussions. 
RR is supported by the Government of Canada through NSERC and by the 
Province of Ontario through MEDT. 
Additional support was provided by the American National Science Foundation 
during the workshop ``Topological Phases and Quantum Computation'' at KITP. 
JH is supported by DTO.
}


\begin{thebibliography}{99}
\bibitem{ShorE}
    P. W. Shor, {\em{Proc. 37th Annual Symp. on the Foundations of Computer Science}}, 56 (IEEE, Los Alamitos, 1996).

\bibitem{TT1}
  E. Knill, R. Laflamme, and W. H. Zurek, Proc. Roy. Soc. London A {\bf{454}}, 365 (1998).

\bibitem{TT2}
  D. Aharonov and M. Ben-Or, {\em{Proc. 29th Annual Symp. on Theory of 
  Computing}}, 176 (ACM, New York, 1997);  D. Aharonov and M. Ben-Or, quant-ph/9906129.

\bibitem{TT3}
    D. Gottesman, Ph.D. thesis, Caltech (1997), quant-ph/9705052.

\bibitem{TT4}
  P. Aliferis, D. Gottesman, and J. Preskill, Quant. Inf. Comp. {\bf{6}}, 97 (2006). 

\bibitem{Kn2}
  E. Knill, Nature {\bf{434}}, 39 (2005).

\bibitem{BT}
  S. Bravyi, Phys. Rev. A {\bf{73}}, 042313 (2006).

\bibitem{BMD}
  H. Bombin and M. A. Delgado, Phys. Rev. Lett. {\bf{97}}, 180501 (2006);
  H. Bombin and M.A. Delgado, quant-ph/0610024. 

\bibitem{SKoh}
  C. M. Dawson and M. A. Nielsen, Quant. Inf. Comp. {\bf{6}}, 81 (2006).

\bibitem{Bon}
  D. Stepanenko and N. E. Bonesteel, Phys. Rev. Lett. {\bf{95}}, 140503 (2005).

\bibitem{Svo}
  K. M. Svore, D. P. DiVincenzo, and B. M. Terhal, quant-ph/0604090.

\bibitem{Ola}
  D. Jaksch {\em{et al.}}, Phys. Rev. Lett. {\bf{82}}, 1975 (1999).

\bibitem{Hen}
  W. K. Hensinger {\em{et al.}}, Appl. Phys. Lett. {\bf{88}}, 034101 (2006).
  
\bibitem{Kit1}
  A. Kitaev, Ann. Phys. {\bf{303}}, 2 (2003).

\bibitem{BK04}
  S. Bravyi and A. Kitaev, Phys. Rev. A {\bf{71}}, 022316 (2005).

\bibitem{RBB03}
  R. Raussendorf, D. E. Browne, and H. J. Briegel, Phys. Rev. A {\bf{68}} (2003). 

\bibitem{RHG}
  R. Raussendorf, J. Harrington, and K. Goyal, Ann. Phys. {\bf{321}}, 2242 (2006). 

\bibitem{DKLP}
  E. Dennis {\em{et al.}}, J. Math. Phys. {\bf{43}}, 4452 (2002).

\bibitem{Ohno}
  T. Ohno {\em{et al.}}, Nucl. Phys. B {\bf{697}}, 462 (2004). 

\bibitem{TSN}
  K. Takeda, T. Sasamoto and H. Nishimori, J. Phys. A {\bf{38}}, 3751 (2005).

\bibitem{AlgoEff}
  J. Edmonds, Can. J. Math {\bf{17}}, 449 (1965).

\bibitem{Harri}
  C. Wang, J. Harrington, and J. Preskill, Ann. Phys. {\bf{303}}, 31 (2003).

\bibitem{Kit2}
  S. Bravyi and A. Kitaev, quant-ph/9811052.

\bibitem{NC00}
  M. A. Nielsen and I. L. Chuang, {\em{Quantum Computation and 
  Quantum Information}} (Cambridge University Press, Cambridge, UK, 2000).

\end{thebibliography}
\end{document}